\documentclass[10pt, twocolumn, notitlepage,prl, superscriptaddress, longbibliography]{revtex4-2}

\usepackage{amsmath}
\usepackage{graphicx}
\usepackage{braket}
\usepackage{lettrine}
\usepackage{xcolor}
\usepackage{hhline}
\usepackage{tabularx}
\usepackage{calc}
\usepackage{soul}
\usepackage{array}

\usepackage{booktabs}
\usepackage{siunitx}
\usepackage{placeins} 

\usepackage[colorlinks=true,linkcolor=blue,citecolor=blue,urlcolor=blue]{hyperref}

\begin{document}

\title{Pure Orbitronics:
Hall Responses and Torques by Orbital Exchange
in Spinless Transition-Metal Dichalcogenides
}

\author{Mahmoud Zeer}
\email{m.zeer@fz-juelich.de}
\affiliation{Peter Gr\"unberg Institute and Institute for Advanced Simulation, Forschungszentrum J\"ulich and JARA, 52425 J\"ulich, Germany}

\author{Dongwook Go}
\affiliation{Department of Physics, Korea University, Seoul, Republic of Korea}

\author{Yuriy Mokrousov}
\email{y.mokrousov@fz-juelich.de}
\affiliation{Peter Gr\"unberg Institute and Institute for Advanced Simulation, Forschungszentrum J\"ulich and JARA, 52425 J\"ulich, Germany}
\affiliation{Institute of Physics, Johannes Gutenberg-University Mainz, 55099 Mainz, Germany}

\begin{abstract}

Recent studies have identified orbital exchange fields as a novel mechanism for controlling electronic and magnetic phenomena through the orbital degree of freedom. In contrast to conventional spin exchange fields, orbital exchange fields couple directly to orbital angular momentum and can emerge from purely orbital interactions or dynamically through couplings to chiral excitations such as phonons and magnons. Motivated by these developments, we investigate the transport and torque responses generated by orbital exchange fields in two-dimensional materials. Using monolayer H-phase MoS$_2$ as a prototypical transition-metal dichalcogenide, we introduce a Zeeman-like orbital exchange field that breaks time-reversal symmetry in orbital space while leaving the spin degree of freedom inactive. We demonstrate that this perturbation induces a sizable intrinsic anomalous Hall conductivity even in the absence of spin magnetism and spin--orbit coupling driven mechanisms. Simultaneously, the orbital Hall response remains robust in the weak-field regime and undergoes substantial reconstruction at larger exchange strengths due to pronounced band modifications. In addition, we show that currents can generate sizable current-induced orbital torques on the orbital magnetization. Our results establish orbital exchange fields as an effective route toward Hall transport and nonequilibrium torque phenomena driven purely by orbital physics, opening new perspectives for orbitronics and orbitally engineered quantum functionalities.
\end{abstract}

\date{\today}
\maketitle

\section{Introduction}

Orbitronics, which exploits the orbital degree of freedom of electrons rather than their spin~\cite{tanaka2008intrinsic,Bernevig2005Orbitronics,bhowal2020intrinsic,cysne2025orbitronics,wang2025orbitronics,chen2024topology,fukami2025challenges,jo2018gigantic}, has recently emerged as a promising alternative to conventional spintronics.
While in conventional spintronics the anomalous Hall effect (AHE)~\cite{nagaosa2010anomalous,xiao2010berry} and current-induced torques~\cite{manchon2009theory}, which emerge as a result of time-reversal symmetry breaking and relativistic effects,  play a central role, the stringent conditions imposed by the strength of spin-orbit coupling (SOC) considerably restrict the classes of materials that can be employed for efficient spintronics applications~\cite{manchon2019current,go2021orbitronics}. 

Unlike spin-based mechanisms that rely on SOC, orbital transport can arise from intrinsic band-structure properties associated with nonrelativistic momentum-space orbital textures, thereby broadening the range of materials capable of hosting non-trivial Hall responses and current-induced torques~\cite{ding2020harnessing,lee2021orbital,hayashi2023observation,go2018intrinsic}. In particular, orbital torque has attracted growing attention as a robust counterpart to conventional spin--orbit torque~\cite{ando2025orbitronics}. 

Indeed, recent advances in orbitronics have established orbital angular momentum as an independent and active degree of freedom in solids that essentially extends the material base for applications. For instance, the orbital Hall effect (OHE) can emerge even in centrosymmetric and nonmagnetic materials, driven purely by interband coherence among multi-orbital states ~\cite{go2018intrinsic}. Furthermore, orbital currents can be converted into spin accumulation through SOC, highlighting orbital physics as a versatile platform for electronic functionalities beyond conventional spin-based paradigms.

In parallel, increasing attention has recently been devoted to the role of orbital exchange fields as a fundamentally new route for controlling magnetic and transport phenomena. Recent theoretical works have demonstrated that orbital exchange interactions can generate purely orbital analogs of chiral magnetic interactions, including orbital Dzyaloshinskii--Moriya interactions (DMI)~\cite{chen2024Dzyaloshinskii}. Moreover, effective orbital exchange fields have been proposed to emerge dynamically through the coupling of electrons to chiral bosonic excitations such as chiral phonons and magnons ~\cite{chaudhary2024giant,juraschek2022giant,luo2023large,wang2024magnetic}. These developments suggest that orbital exchange fields are not merely theoretical constructs, but can arise naturally in realistic nonequilibrium quantum materials. This  strongly motivates the exploration of orbital exchange fields as a distinct platform for transport  phenomena beyond conventional spin magnetism. Remarkably, it was recently shown that the coupling between nonequilibrium orbital angular momentum and orbital exchange fields can produce torques acting on the spin degree of freedom, opening a new pathway toward orbitally driven spin manipulation~\cite{lee2026orbital,schmitt2026orbital}. 

\begin{figure}[t!]
\centering
\includegraphics[width=0.48\textwidth]{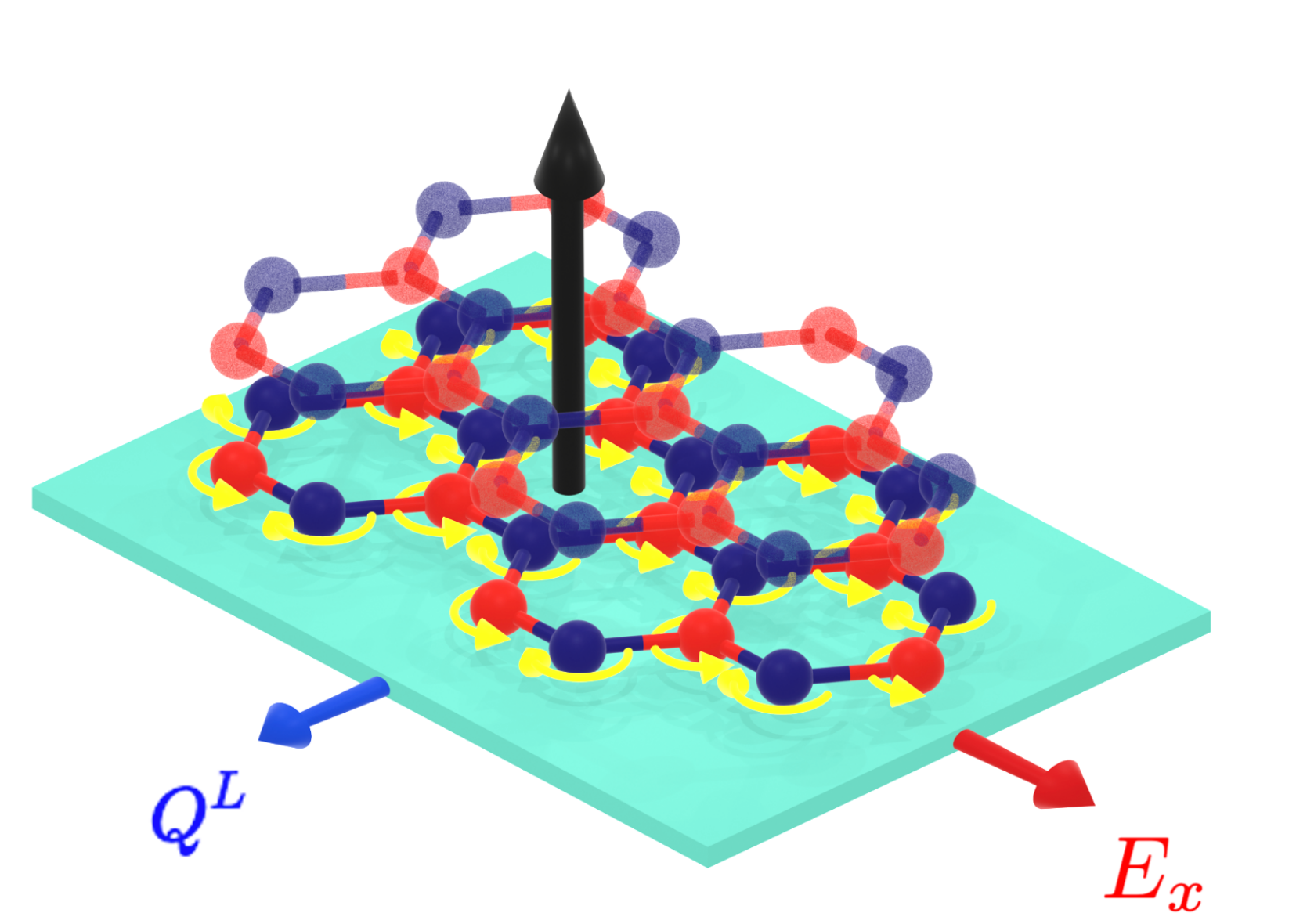}
\caption{
Crystal structure and schematic illustration of orbital exchange in TMDs.
Monolayer H-MoS$_2$ with the applied orbital Zeeman--type exchange field $\boldsymbol{B}_{\mathrm{orb}}$ coupling to orbital angular momentum $\hat{\boldsymbol{L}}$. The applied in-plane electric field $E_x$ generates transverse charge and orbital responses. The underlying network of yellow arrows depicts the realization of  orbital exchange field generation by a chiral phonon.
}
\label{fig:Structure}
\end{figure}

Despite this progress, a fundamental question remains unresolved: can the whole range of transport effects normally associated with relativistic magnets -- i.e. anomalous Hall effect, orbital Hall effect and current-induced torques -- arise purely from orbital degrees of freedom without relying on SOC or spin magnetism? In this work, we demonstrate that such a mechanism is indeed possible. Using monolayer H-phase MoS$_2$ as a prototypical two-dimensional transition-metal dichalcogenide (TMD), we introduce a Zeeman-like orbital exchange field that couples directly to the orbital angular momentum operator Fig~\ref{fig:Structure}. This perturbation breaks time-reversal symmetry in orbital space while leaving the spin degree of freedom inactive. We show that even a weak orbital exchange field induces a sizable intrinsic anomalous Hall conductivity (AHC), while the orbital Hall conductivity (OHC) remains remarkably robust in the weak-field regime. As the exchange strength increases, pronounced band reconstruction qualitatively modifies both Hall responses. Moreover, the orbital exchange field generates substantial current-induced orbital torques. Our results establish orbital exchange fields as a powerful mechanism for engineering Hall responses and nonequilibrium torques in the absence of conventional spin magnetism and SOC-driven effects, opening new directions for orbitronic functionalities and orbitally controlled quantum devices.

\begin{figure*}[t!]
\includegraphics[width=0.95\textwidth]{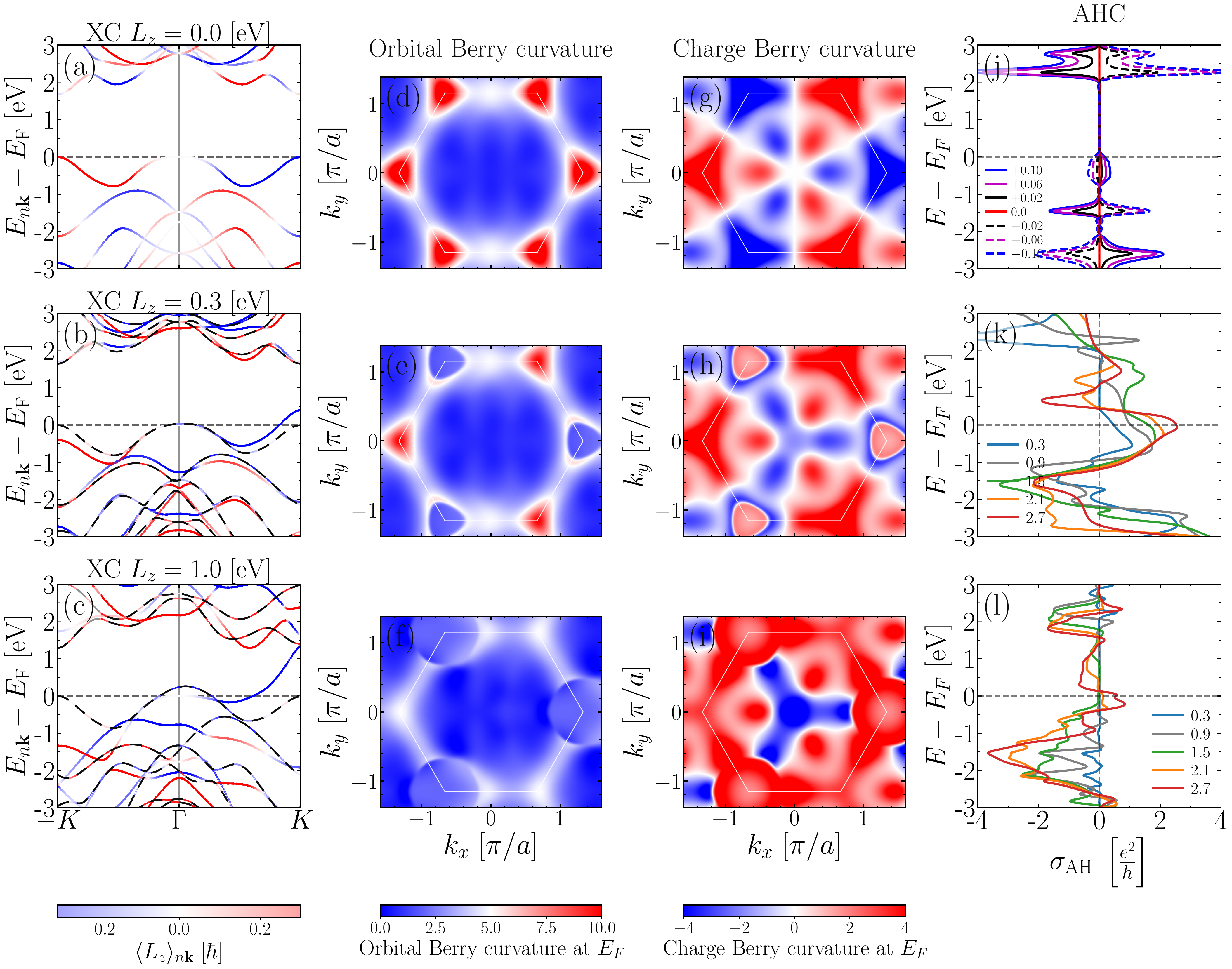}
\caption{
\textbf{Orbital-exchange--induced anomalous Hall effect.}
(a--c) Orbitally-resolved band structure colored by the expectation value of the out-of-plane orbital angular momentum, $\langle L_z \rangle_{n\mathbf{k}}$, for orbital exchange field applied along the $z$ axis with the strength of (a) 0, (b) 0.3, and (c) 1.0~eV. The color scale denotes $\langle L_z \rangle_{n\mathbf{k}}$ in units of $\hbar$. Dashed black lines show the corresponding band structures for an exchange field of the same magnitude applied along the $y$ direction.
(d--f) Orbital Berry curvature evaluated at the Fermi energy for orbital exchange strength of (d) 0, (e) 0.3, and (f) 1.0~eV.
(g--i) Charge Berry curvature at the Fermi energy for the same exchange strength as in (d--f). Color scales for the Berry curvatures are shown with a color bar.
(j) Energy-resolved anomalous Hall conductivity (AHC) for small positive and negative orbital exchange fields, highlighting the time-odd nature of the response. The AHC vanishes in the absence of orbital exchange. 
(k,l) AHC as a function of Fermi energy for increasing orbital exchange strengths applied along (k) $z$-axis  and (l) $y$-axis, revealing a pronounced directional anisotropy of the orbital-exchange--driven AHE. The values of the orbital exchange field are shown in the legends in units of eV.}
\label{fig:bands_and_AHC_Lz_Ly}
\end{figure*}

\section{Results and discussion}

\subsection{Orbital-exchange-induced anomalous Hall effect}

Starting from the Hamiltonian of MoS$_2$ (see Methods), we introduce an  orbital Zeeman-type of exchange field by adding the following term to the single-particle Hamiltonian:
\begin{equation}
\hat{H}_{\mathrm{XC}} = \boldsymbol{B}_{\mathrm{orb}} \cdot \hat{\boldsymbol{L}},
\end{equation}
where $\hat{\boldsymbol{L}}$ is the atomic orbital angular momentum operator and $\boldsymbol{B}_{\mathrm{orb}}$ denotes the orbital exchange field, which we apply here applied along  $y$ or $z$ directions. In our approach, all quantities are expressed in terms of matrix elements evaluated in the basis of maximally-localized Wannier functions, see section Methods.
The anomalous Hall conductivity (AHC) is calculated within the Kubo linear-response formalism by invoking the expression
\begin{equation}
\sigma_{xy}^{\mathrm{AH}} = -\frac{e^{2}}{\hbar} \sum_{n} \int_{\mathrm{BZ}} \frac{d^{2}k}{(2\pi)^{2}}
f_{n\boldsymbol{k}} \, \Omega_{xy}^{n}(\boldsymbol{k}),
\end{equation}
\begin{equation}
\Omega_{xy}^{n}(\boldsymbol{k})=2\hbar^{2} \sum _{n\neq m}
\mathrm{Im}
\left[
\frac{\left<u_{n\boldsymbol{k}} \left|v_{y}\right|u_{m\boldsymbol{k}}\right>\left< u_{m\boldsymbol{k}}\left|v _{x}\right|u_{n\boldsymbol{k}}\right>}{(E_{n\boldsymbol{k}}-E_{m\boldsymbol{k}}+i\eta )^2}
\right],
\end{equation}
where $f_{n\boldsymbol{k}}$ is the Fermi--Dirac distribution and $\Omega_{xy}^{n}(\boldsymbol{k})$ the Berry curvature of band $n$. Here, $\boldsymbol{k}$ is the crystal momentum, $|u_{n\boldsymbol{k}}\rangle$ the cell-periodic Bloch states, $v_{x,y}$ the velocity operators, $E_{n\boldsymbol{k}}$ the band energies, and $\eta$ a small broadening parameter; the sum over $m \neq n$ accounts for interband contributions.
The OHC was obtained analogously by replacing the charge current operator with the orbital current operator
$\hat{J}_{x}^{L_{\alpha}} = \frac{1}{2} \left\{ \hat{v}_{x}, \hat{L}_{\alpha} \right\}.$

In Fig.~\ref{fig:bands_and_AHC_Lz_Ly} (a--l) we show the evolution of the band structure, Berry curvature, and anomalous Hall conductivity (AHC) under applied orbital exchange field.
In the absence of spin--orbit coupling  and orbital exchange, monolayer H-MoS$_2$ preserves time-reversal symmetry and  exhibits vanishing  integrated Berry curvature  [Fig.~\ref{fig:bands_and_AHC_Lz_Ly}(g)]. It is well-known that although valley-resolved Berry curvature hot spots appear near K and K$^\prime$, their contributions cancel exactly  resulting in zero intrinsic AHC, as shown in red in Fig.~\ref{fig:bands_and_AHC_Lz_Ly}(g,j).
Applying an orbital exchange field breaks time-reversal symmetry while leaving the spin degrees of freedom inactive without spin-orbit coupling, the effect of which we neglect in this study. Consequently, the spin degree of freedom is frozen out and the band always remain degenerate in spin. 

\begin{figure*}[!t]
\centering
\includegraphics[width=0.75\textwidth]{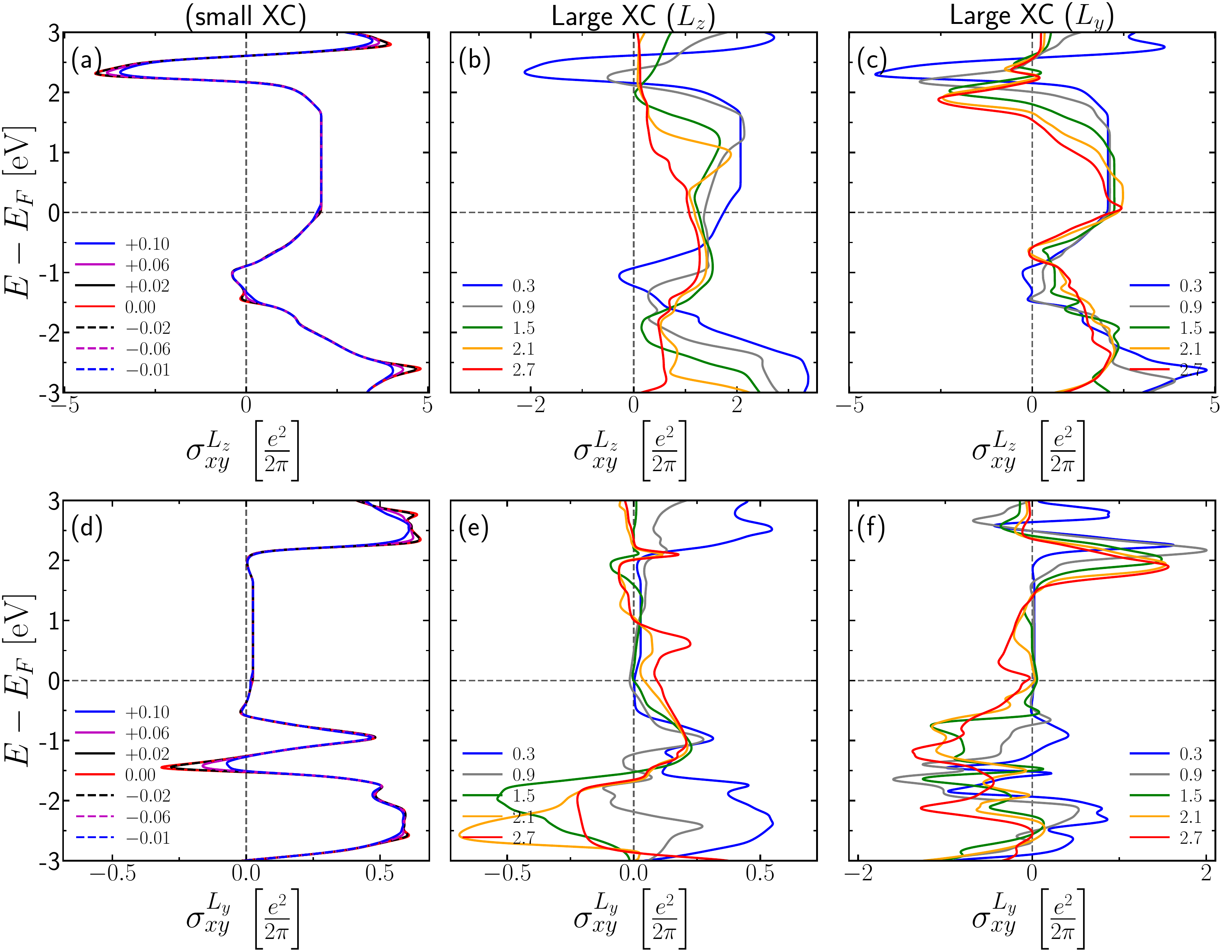}
\caption{
\textbf{$L_z$- and $L_y$-polarized orbital Hall conductivity (OHC) as a function of the orbital exchange field.}
(a--c) OHC with orbital polarization of the current along the $z$ direction, $\sigma_{xy}^{L_z}$.
(a) OHC as a function of Fermi energy for small positive and negative orbital exchange fields ranging from $+0.01$ to $-0.10$~eV, showing negligible variations and confirming that the OHC is a time-even (T-even) response whose sign is insensitive to the direction of the orbital exchange field.
(b) OHC for increasing strengths of orbital exchange applied along the $z$ direction. The OHC remains comparable to the zero-exchange case up to $\sim 0.6$~eV, where the underlying band structure is only weakly perturbed; for larger exchange strengths, pronounced band-structure reconstructions lead to qualitatively different OHC behavior.
(c) Same as (b), but for orbital exchange applied along the $y$ direction, highlighting the directional dependence of orbital-exchange-induced modifications to the OHC.
(d--f) Same as (a--c), but for OHC component $\sigma_{xy}^{L_y}$.
}
\label{fig:Orbital_exchange_OHC}
\end{figure*}

In this work, we consider two regimes of orbital exchange: weak and strong. Note that the band structure remains comparable to that without orbital exchange up to $\sim 0.7$~eV; beyond this value, the band structure changes dramatically. For the weak regime, shown in Fig.~\ref{fig:bands_and_AHC_Lz_Ly}(j), even for small exchange strengths ($\lvert B_{\mathrm{orb}} \rvert \lesssim 0.1$~eV), we observe lifting of valley degeneracy (see Fig.~ S1 (c)), which leads to redistribution of orbital character $\langle L_z \rangle_{n\mathbf{k}}$ near $K/K^\prime$, the emergence of a finite Berry-curvature imbalance, and emergence of finite AHC.
As shown in Fig.~\ref{fig:bands_and_AHC_Lz_Ly}(j), the AHC changes sign upon reversing the exchange field, confirming its time-odd character (see Fig.~S1 (c) and (d)). The response is approximately linear in $B_{\mathrm{orb}}$ in the weak-field regime; see Table~ S1, consistent with first-order perturbation theory in the time-reversal--breaking field.

We can physically understand the mechanism as follows. When the orbital exchange field is applied out of plane, the states with opposite orbital polarization at K and K$^{\prime}$ move in opposite direction in energy. This modifies the interband matrix elements of the velocity operators, changes the position of the states in energy and their occupation  in each valley in an asymmetric way, which in turn  reshapes the Berry-curvature distribution in momentum space  and the Brillouin-zone integral acquires a finite value. 
This effect is clearly observed when comparing the Berry curvature in Fig.~\ref{fig:bands_and_AHC_Lz_Ly}(g) (without orbital exchange) to Fig.~\ref{fig:bands_and_AHC_Lz_Ly}(h) (with an orbital exchange of 0.3~eV). Since without the field each valley contributes with opposite AHC, when the orbital field changes sign, the reversal of the integrated AHC follows.

We now turn to the regime of strong orbital exchange fields applied along $z$, which we define as the region of exchange strength ranging from 0.3~eV up to 2.7~eV,~Fig.~\ref{fig:bands_and_AHC_Lz_Ly}(k).
In this regime, substantial band reconstruction occurs [Fig.~\ref{fig:bands_and_AHC_Lz_Ly}(b,c)]. Avoided crossings shift significantly in energy, and the orbital polarization develops a pronounced momentum dependence. Consequently, the Berry-curvature hot spots migrate in momentum space, and additional curvature emerges from newly formed avoided crossings.
The energy-resolved AHC exhibits a linear in energy behavior at the Fermi level, with sign reversals occurring as a function of the Fermi energy.

We next clarify the directional dependence of the AHC  by applying the orbital exchange field along $y$. By comparing the case of $\boldsymbol{B}_{\mathrm{orb}}\Vert z$ [Fig.~\ref{fig:bands_and_AHC_Lz_Ly}(k)] to the case of $\boldsymbol{B}_{\mathrm{orb}}\Vert y$ [Fig.~\ref{fig:bands_and_AHC_Lz_Ly}(l)], 
we notice a pronounced anisotropy with respect to the orbital exchange field direction.
This anisotropy originates from the distinct symmetry-breaking patterns associated with the two field orientations. An orbital exchange field along $z$ preserves in-plane rotational symmetry while breaking time-reversal symmetry. In contrast, for $\mathbf{B}_{\mathrm{orb}}$ along $y$  the in-plane rotational symmetry is lowered and the valley coupling is modified in a qualitatively different manner. 

The corresponding band structures for orbital exchange applied along $y$  are shown as dashed lines in Figs.~\ref{fig:bands_and_AHC_Lz_Ly}(a,b). From these plots we observe that the effect of the in-plane orbital exchange field on the bands is minimal around $K/K^\prime$, where the states are strongly polarized along $z$. On the other hand,  these are the bands around the $\Gamma$-point that are effected strongest by the in-plane field, which can be attributed to the stronger impact of the orbital Rashba effect in that part of the Brillouin zone, leading to a stronger in-plane orbital polarization of the states. The drastic difference in the evolution in the bands with the strength of the field for the two cases results in a strongly anisotropic AHC, which is much more suppressed around the Fermi energy for the in-plane field.

\subsection{Orbital-exchange-induced OHC}

The orbital Hall effect in MoS$_2$, as a representative two-dimensional transition-metal dichalcogenide, has attracted considerable attention in recent years~\cite{cysne2021disentangling}. This motivated us to adopt monolayer MoS$_2$ as a prototypical platform to investigate orbital transport and current-induced orbital torque at the first place. In the absence of orbital exchange, our calculated band structure, orbital Berry curvature, and orbital Hall conductivity are in excellent agreement with previous reports~\cite{bhowal2020intrinsic,cysne2021disentangling,canonico2020orbital}, thereby validating our computational framework.

Following the same analysis strategy as in the previous section, we now examine the evolution of the orbital Hall conductivity under finite orbital exchange, presenting the results in Fig.~\ref{fig:Orbital_exchange_OHC}. The first row of Fig.~\ref{fig:Orbital_exchange_OHC} displays the $L_z$-polarized OHC, while the second row shows the corresponding $L_y$-polarized response. Panels (a) and (d) present the weak-exchange regime, where only minor modifications are observed. For larger exchange strengths applied along the $z$ direction, the results are shown in panels (b) and (c), whereas panels (e) and (f) correspond to strong exchange applied along the $y$ direction.

A key qualitative distinction between the OHC and the AHC, valid in both regimes of the orbital field strength, is that while the AHC [Fig.~\ref{fig:bands_and_AHC_Lz_Ly}(j)] exhibits a time-odd character, the OHC [Fig.~\ref{fig:Orbital_exchange_OHC}(a,d); see also Fig.~ S1(a,b)] remains manifestly time-even and it is therefore invariant under exchange-field reversal. Throughout the weak-exchange regime, notably, the magnitude of the orbital Hall response remains largely robust, for both flavors of the orbital current polarization. In this regime, the only visible changes in the OHC can be observed around $-2.5$, $-1.5$ and $+2.5$\,eV. These correspond to the positions in energy where OHC exhibits extrema and where the impact of the orbital field on the bands around the $\Gamma$-point is most drastic in terms of band degeneracy lifting,~Fig.~\ref{fig:Orbital_exchange_OHC}(a,d).
We also observe that among the two types of orbital polarization, the $L_z$-OHC exhibits a much larger magnitude around $E_F$, owing to the strong orbital out-of-plane polarization of the states around $E_F$ at $K$ and $K^\prime$, which results in very large contributions to the orbital Berry curvature from the valleys, see Fig.~\ref{fig:bands_and_AHC_Lz_Ly}(d). In case of $\sigma_{xy}^{L_y}$, however, the diagonal matrix elements of $L_y$ are suppressed around the Fermi energy, while the off-diagonal matrix elements of $L_y$ contribute to the OHC via transitions among bands which are well separated in energy. Notably, $\sigma_{xy}^{L_y}$ is practically absent within the gap.

We now turn to the strong-exchange regime and analyze the evolution of the OHC as a function of exchange field applied along $z$, focusing in particular at the region of the gap, Fig.~\ref{fig:Orbital_exchange_OHC}(b). We observe that the magnitude of $\sigma_{xy}^{L_z}$ is consistently suppressed here as the strength of the exchange field increases. This can be understood by following the evolution of charge and orbital Berry curvature in $k$-space, Fig.~\ref{fig:bands_and_AHC_Lz_Ly}. Upon increase in $B_z$, the valence band at $K$ becomes partially unoccupied, which result in a suppression of the charge Berry curvature, and correspondingly, in orbital Berry curvature. On the other hand, at $K^\prime$, the top valence band moves well below the Fermi level with increasing orbital field, where it mixes with the valence bands of different orbital symmetry, thus triggering suppression of $L_z$-orbital Berry curvature as well.  

Concerning $\sigma_{xy}^{L_y}$, at intermediate values of the orbital exchange field  the OHC switches from a small positive to a small negative value. This can be explained by the fact that, since $L_y$ conductivity relies predominantly on interband matrix elements in this regime, the contribution at $K$ becomes suppressed as the top valence band moves above $E_F$, while the orbital properties at $M$ are only weakly effected by the orbital field along $z$, which tips the balance toward the negative values of the OHC, see Fig.~S3. Upon further increase in the magnitude of the exchange field, the bands around the Fermi energy are strongly reorganized in the sense that the states with $d_{xz,yz}$ and $p_{x,y}$ symmetry become available for forming linear combination with $d_{xy,x^2-y^2}$ states in the same region of energy, which significantly increases $\sigma_{xy}^{L_y}$ at the Fermi energy. Thus we observe a ``transfer" of $L_z$-OHC magnitude into $\sigma_{xy}^{L_y}$.

The same effect can be observed also in the Fermi energy region for an exchange field applied along $y$. In this case, as evident from Fig.~\ref{fig:bands_and_AHC_Lz_Ly}, the changes in relative positions of the bands with $B_y$ are much smaller, which results in much smoother evolution of the orbital Berry curvature distribution with the exchange field ~Fig. S3. We also observe that the changes in the values of both OHCs are primarily associated with contributions to the orbital Berry curvature around $M$-points.
The pronounced modification of the orbital Hall conductivity near the $M$-point originates from enhanced orbital hybridization  in this region of the Brillouin zone. The orbital exchange field along $y$ efficiently mixes the competing Mo $d$-orbital manifolds near $M$, where saddle-point-like dispersion and symmetry-protected crossings strongly enhance the orbital Berry-curvature response. In contrast to the valley states near $K$, which are dominated by relatively well-defined and robust $d_{x^2-y^2}\pm i d_{xy}$ character, the electronic structure near $M$ involves substantial admixture of multiple orbital channels, including $d_{xz}$, $d_{yz}$, $d_{xy}$, and $d_{x^2-y^2}$ states of Mo and sulfur $p$-states. The orbital exchange field along $y$ lowers the crystalline symmetry by breaking mirror and rotational symmetries that otherwise constrain the orbital texture, thereby lifting degeneracies and enhancing inter-manifold coupling. As a result, the distribution of orbital polarization and associated orbital Berry curvature near $M$ become strongly susceptible to the magnitude of the exchange field, leading to the pronounced changes in the orbital Hall conductivity.

\begin{figure}[t!]
\centering
\includegraphics[width=0.48\textwidth]{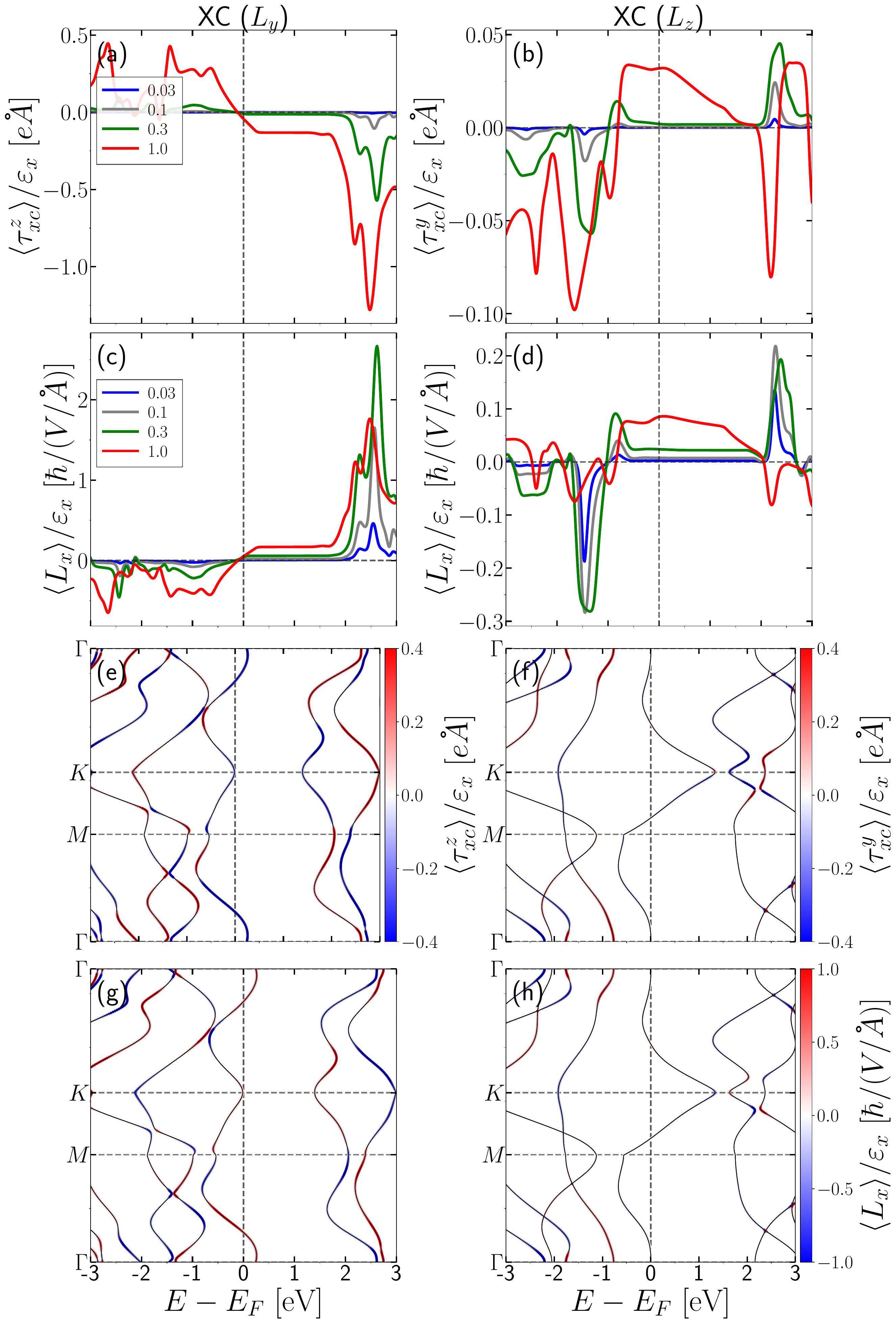}
\caption{{
{{\textbf{ Orbital torque and current-induced orbital polarization generated by orbital exchange in monolayer MoS$_2$.}} 
(a,b) Band-filling dependence of the orbital-exchange torque normalized to the applied electric field $E_x$, for orbital exchange fields applied along the (a) $y$ and (b) $z$ directions.
(c,d) Corresponding band-filling dependence of the current-induced orbital angular momentum $\langle L_x \rangle$, normalized to $E_x$, for orbital exchange applied along the (c) $y$ and (d) $z$ directions.
(e,f) Orbital-exchange torque normalized to $E_x$ for a fixed exchange strength of $1.0$~eV applied along the (e) $y$ and (f) $z$ directions.
(g,h) Corresponding current-induced orbital angular momentum $\langle L_x \rangle$ normalized to $E_x$ for the same exchange strength ($1.0$~eV).
}}}
\label{fig:Torque_and_Orbitat_moment}
\end{figure}

\subsection{Orbital-exchange-induced orbital torque and orbital accumulation}

In the conventional picture, spin–orbital, or, orbital torque  ultimately arises from from the exchange of angular momentum between spin and orbital degrees of freedom mediated by magnetic exchange interaction and SOC~\cite{manchon2019current,sinova2015spin}. An electric field generates a nonequilibrium orbital polarization (e.g., via the orbital Hall~\cite{go2020orbital,choi2023observation} or orbital Edelstein effects~\cite{fukunaga2023orbital}), which is converted into a torque on the magnetic order through exchange interaction~\cite{liu2024harnessing,ding2024orbital,ding2024mitigation}.
Here, we examine whether an analogous mechanism can arise purely from orbital degrees of freedom in our system. 
We follow the idea that when an electric field is applied to our system subject to an orbital exchange field, this will generate  non-equilibrium orbital polarization. At the same time time, following the continuity equation for the density of orbital angular momentum~\cite{go2020orbital}, the dynamics of OAM will be governed by a competition among the crystal field torque and the torque mediated by a commutator between the operators of OAM and orbital exchange field. In fact, such {\it orbital exchange torque} (or simply orbital torque in this work) can be seen as an analogy of the spin-orbital torque computed when replacing the operator of spin with identity in spin-orbit interaction potential (i.e. working within a manifold of spin-pure states):
\begin{equation}
\hat{\boldsymbol{\tau}} = \frac{i}{\hbar} [\hat{H}_{\rm XC}, \hat{\boldsymbol{L}}] = \frac{iB_{\rm orb}}{\hbar} [L_{z(y)}, \hat{\boldsymbol{L}}],
\end{equation}
when the orbital exchange field points along $z$ ($y$). Correspondingly, assuming that in the basis of Wannier functions the commutation relations between the OAM operators are approximately satisfied, we come to the conclusion that for the case of $z$-orbital field the torque is determined either by non-equilibrium $\delta L_x$ or $\delta L_y$, while in the case of $y$-field it is the $\delta L_x$ or $\delta L_z$ which determine the orbital torque behavior. 

From the viewpoint of symmetry, when the electric field is applied along the $x$-axis, among relevant for the torque components only $\delta L_x$ component of induced OAM is allowed. Indeed, in case of the orbital field along $z$, applying $\mathcal{M}_y$ in combination with time-reversal $\mathcal{T}$ keeps the crystal Hamiltonian intact and preserves  the directions of the orbital exchange field and electric field, while flipping the sign of $\delta L_y$, implying $\delta L_y=0$. Analogously, for the case of the orbital exchange field along $y$, it is the $\mathcal{M}_z\mathcal{T}$ operation which flips the sign of $\delta L_z$ while keeping the Hamiltonian and electric field intact, implying $\delta L_z=0$. Correspondingly, we compute the current-induced orbital accumulation along $x$, presenting the band-filling results in Fig.~\ref{fig:Torque_and_Orbitat_moment}(c,d,g,h) in direct comparison to the orbital torque Fig.~\ref{fig:Torque_and_Orbitat_moment}(a,b,e,f) for the orbital exchange field along $z$ (right panel) and $y$ (left panel).

A clear correlation can be observed  between the induced orbital polarization $\langle L_\alpha \rangle$ and the resulting exchange torque. For both directions of the orbital exchange field, the energy dependence of the torque closely follows that of the current-induced orbital angular momentum, directly confirming that the torque originates from the coupling between nonequilibrium orbital accumulation and the orbital exchange field. 
As the magnitude of the exchange field increases, both the induced orbital polarization and the corresponding torque grow systematically, with pronounced features near band edges where interband coherence effects are strongest -- which is also the case for orbital Hall conductivities. Notably, a strong anisotropy is observed: an exchange field applied along $y$ mediates substantially larger torques than in the case of $z$-exchange field, reflecting the symmetry-selective efficiency of orbital-dynamics in the two-dimensional lattices with lowered symmetry. 

The anisotropy of the torque with respect to the exchange field direction is also clearly visible upon inspection of the band-resolved data in Fig.~\ref{fig:Torque_and_Orbitat_moment}(e-h): the much stronger response of $\delta L_x$ and the torque for $y$-field is a consequence of a collective effect, when more bands participate in the torque formation -- this is in contrast to the case of the $z$-field which mediates the torque predominantly around selected band crossings and band edges. This can be traced back to the fact that it is fundamentally easier to achieve an in-plane non-equilibrium orbital polarization when the states are already somewhat in-plane polarized by the $y$-field, while in the case of the out of plane orbital field the formation of the in-plane moment by the current has to overcome the tendency of the states to be orbitally polarized along $z$ in large parts of the Brillouin zone around the Fermi energy -- an effect further enhanced by the applied exchange field along $z$. This ultimately crystal field effect finds its manifestation also in the behavior of the orbital Hall conductivities as discussed above, i.e. it is easier to achieve an in-plane polarization of the orbital Hall current for the orbital exchange field in-plane rather than for the out of plane field, Fig.~\ref{fig:Orbital_exchange_OHC}(e,f).

\section{Discussion}

In this work, we have demonstrated that a Zeeman-type orbital exchange field induces a sizable intrinsic anomalous Hall effect in monolayer MoS$_2$ in the complete absence of spin--orbit coupling. While the orbital Hall effect remains remarkably robust in the weak-field regime, strong orbital exchange leads to pronounced band-structure reconstruction and qualitative changes in both charge and orbital transport responses. Furthermore, we showed that orbital exchange generates substantial current-induced orbital torques and nonequilibrium orbital polarization. Taken together, these findings establish orbital exchange as an efficient and versatile mechanism for engineering Hall responses and torque phenomena in nonmagnetic and spin--orbit-free materials, significantly broadening the scope of orbitronics beyond conventional spin-based paradigms.

Our results furthermore establish a purely orbital analogue of spin-orbit torque. In contrast to conventional spin-orbit torque mechanisms, which rely on spin-orbit coupling and magnetic moments, the torque uncovered here arises entirely from orbital angular momentum transport coupled to orbital exchange. This demonstrates that current-induced torque is not intrinsically a spin phenomenon, but can emerge more generally from angular-momentum transport in the presence of exchange kind of coupling to a time-reversal breaking perturbation. The present findings therefore highlight orbital exchange as a distinct route toward nonequilibrium control of quantum degrees of freedom, complementing established spin-based mechanisms and extending the range of materials and interactions that can host torque phenomena.

Although the orbital exchange field introduced in this work is implemented as a model perturbation, it provides a clear and controlled framework for isolating purely orbital contributions to transverse transport and torque phenomena, independent of spin--orbit coupling. More importantly, such an effective field is not merely conceptual. Orbital polarization fields may in principle be engineered through proximity effects, strain-induced orbital splitting, or coupling to materials and excitations with strong orbital character. Recent studies have further suggested that effective orbital exchange fields can emerge dynamically through coupling to chiral bosonic excitations, including chiral phonons and magnons. From this perspective, the orbital-exchange-induced Hall responses and torques reported here can be viewed as generic consequences of {\it orbital polarization fields}, irrespective of whether their origin is electronic, magnetic, or bosonic. Our results therefore identify orbital exchange not only as a theoretical probe, but also as a realistic design principle for realizing spin--orbit-free Hall responses and orbital torques in low-dimensional systems.

This broader viewpoint suggests that the conceptual framework developed here may extend beyond electronic orbitronics into systems hosting orbital-like polarization degrees of freedom. In particular, the strong coupling between nonequilibrium orbital accumulation and orbital exchange identified in our calculations points to possible interactions between charge currents and angular-momentum-carrying collective modes. Circularly polarized phonons and photons possess polarization states that can be viewed as analogues of orbital polarization, raising the possibility that orbital exchange may provide a route for electrical control of chiral phonon polarization and related dynamical phenomena. More generally, the current-induced orbital torque uncovered here can be interpreted as a generic angular-momentum-transfer mechanism mediated by orbital exchange, potentially relevant to hybrid electronic, phononic, and photonic platforms. Such extensions would establish orbital exchange as a unifying concept linking electronic transport, lattice dynamics, and light--matter interactions, opening new opportunities for orbitally controlled functionalities beyond conventional spintronics and magnetism.

\section{Method}

Density functional theory (DFT) calculations for monolayer MoS$_2$ were performed using the full-potential linearized augmented plane-wave (FLAPW) method as implemented in the \textsc{FLEUR} code~\cite{fleur}. The exchange–correlation potential was treated within the Perdew–Burke–Ernzerhof (PBE) generalized gradient approximation~\cite{perdew1996generalized}. A $16 \times 16$ $k$-point mesh was used for Brillouin-zone sampling. Spin–orbit coupling was neglected throughout. The lattice constant of MoS$_2$ is 3.18~\AA, and distance between the planes of Mo and S atoms along the z-axis is 1.56~\AA.

Twenty-two maximally localized Wannier functions (MLWFs) were constructed from Mo-$d$ and S-$p$ orbitals, which dominate the low-energy electronic structure. The frozen energy window extended up to 3~eV above the Fermi level. The Wannier-interpolated band structure was validated against the original FLAPW bands~\cite{Freimuth2008} and independent constructions using \textsc{Wannier90}~\cite{MOSTOFI20142309}. All response functions were evaluated using a dense $500 \times 500$ interpolated $k$-point mesh.

Current-induced orbital torque and nonequilibrium orbital polarization were evaluated within linear response. 

All transport, torque, and orbital polarization quantities are reported per unit electric field, with the electric field applied along the $x$ direction.

\section*{Acknowledgements}
\begin{acknowledgements}
This work was supported by the Federal Ministry of Education and Research of Germany in the framework of the Palestinian-German Science Bridge (BMBF grant number 01DH16027). We also gratefully acknowledge financial support by the Deutsche Forschungsgemeinschaft (DFG, German Research Foundation) $-$ TRR 288 $-$ 422213477 (project B06). This work was supported by the EIC Pathfinder OPEN grant 101129641 “OBELIX”. We also gratefully acknowledge the J\"ulich Supercomputing Centre and RWTH Aachen University for providing computational resources under projects jiff40 and jara0062.
\end{acknowledgements}

\section*{Author Contributions}
Y.M., D.G. and M.Z. initiated and designed the project idea, and Y.M. supervised the project. M.Z. performed the simulations and analysis under the supervision of Y.M. All authors contributed to the discussion and interpretation of the results. M.Z. wrote the draft of the manuscript, and all co-authors contributed to its revision.

\section*{Data Availability}

The data that support the findings of this study are available from the corresponding author upon reasonable request.

\vspace{0.5\baselineskip}
\textbf{Competing Interests}

The authors declare no competing interests.

\bibliography{Orbital_Exchange.bib}

\end{document}